# Light amplification in semiconductor-superconductor structures

Raja Marjieh, Evyatar Sabag and Alex Hayat,

*Department of Electrical Engineering, Technion, Haifa 32000, Israel*

**Abstract.** We study a new effect of Cooper-pair-based two-photon gain in semiconductor-superconductor structures, showing broadband enhancement of ultrafast two-photon amplification. We further show that with the superconducting enhancement, at moderately high seed intensities the two-photon gain contribution approaches that of the one-photon gain. A full quantum-optical model of singly- and fully-stimulated two-photon emission is developed. Our results provide new insights on nonlinear light-matter interaction in the superconducting state, including the possibility of coherent control in two-photon semiconductor-superconductor devices. The theoretically-demonstrated effects can have important implications in optoelectronics and in coherent-control applications.

## 1. Introduction

Superconducting optoelectronics, based on superconductivity in semiconductors is a rapidly growing field of research [1, 2, 3, 4, 5, 6, 7, 8]. One of the many new fascinating phenomena related to superconductivity induced in semiconductors by the proximity effect, is enhanced light emission. In particular – enhanced Cooper-pair-based two-photon emission (TPE). Without superconducting enhancement, TPE has been studied as the basis for an alternative form of quantum oscillators [9,10,11]. Such TPE sources exhibit a rich spectrum of both classical and quantum phenomena including bi-stability and giant pulse generation, as well as completely quantum processes like squeezing [12]. In semiconductors, TPE was demonstrated recently [13,14,15,16] with potential applications of spontaneous TPE in quantum technologies [17,18,19], as well as novel light amplification in electrically-pumped devices based on singly-stimulated TPE [12] and fully-stimulated TPE [20, 21]. This two-photon gain (TPG) allows ultrafast pulse compression [22], and can be further enhanced in the nondegenerate case [23] . Superconducting proximity effect has been shown lately to significantly enhance photon pair emission from semiconductor light-emitting diodes [1] so that the resulting spontaneous emission has a significant contribution of Cooper-pair based TPE (Fig. 1 a). This is a strong enhancement of several orders of magnitude, compared to spontaneous TPE in semiconductors without superconductivity [12]. It suggests that other multi-photon processes could benefit from similar enhancement including the nonlinear effect of semiconductor TPG [15] with promising applications in ultrafast electrically-pumped devices.

Here we propose and analyze theoretically a new effect of enhanced light amplification in electrically-driven semiconductor-superconductor structures, including Cooper-pair based TPG in a superconducting proximity region of the semiconductor. Conduction band Cooper-paired electrons in the proximity region form a many-body Bardeen-Cooper-Schrieffer (BCS) state [24]. This, in turn, yields enhanced emission through recombination of conduction-band Cooper pair electrons with valence-band holes [25]. The core idea of our proposal is to couple a superconducting BCS state to a structure capable of light amplification (a semiconductor) in order to achieve enhanced two-photon gain.

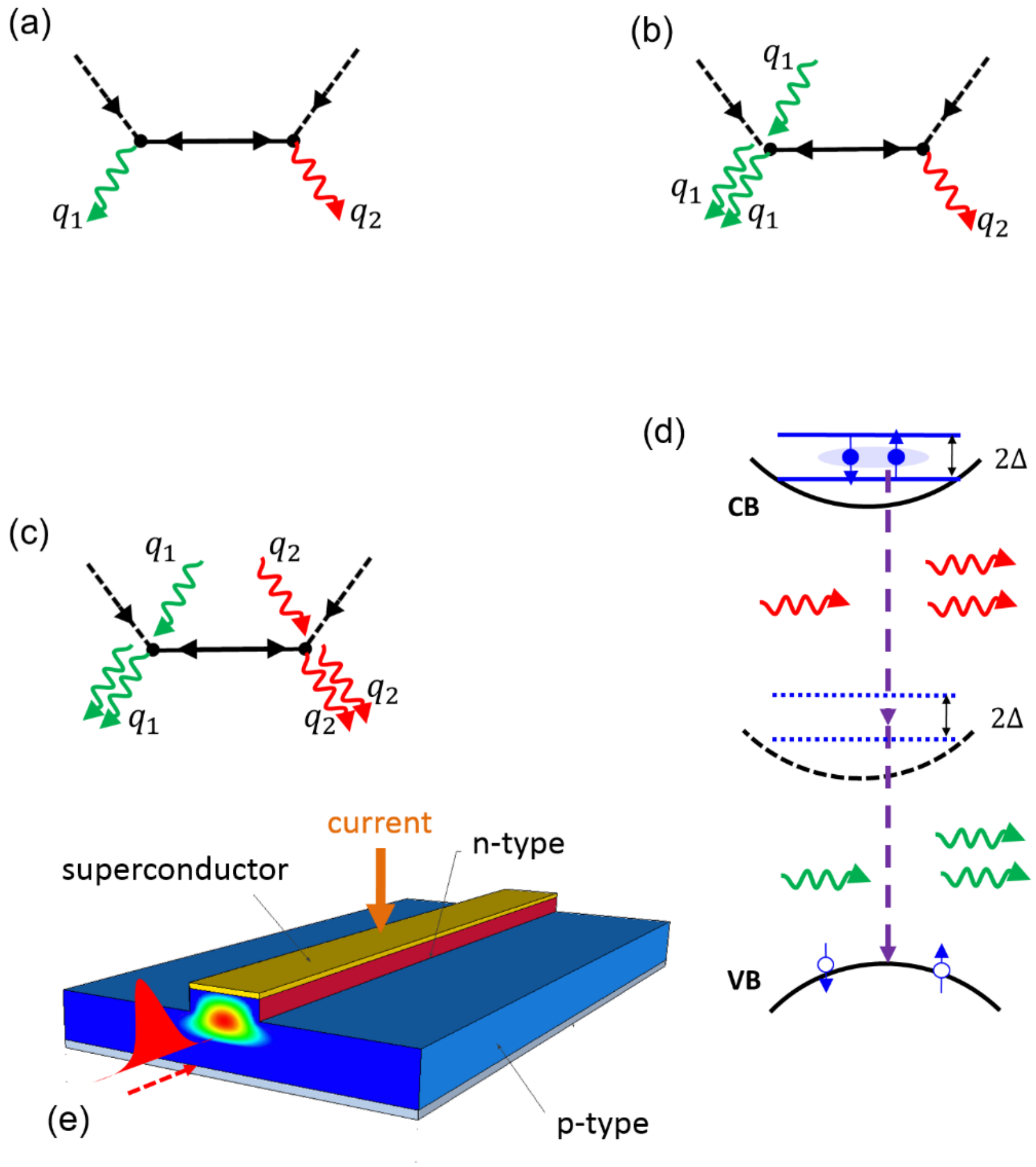

**Figure 1.** Feynman diagrams of the Cooper-pair based emission processes. The solid lines indicate electrons, the dashed lines indicate holes, and the wavy lines indicate photons. (a). Spontaneous TPE. (b) Singly-stimulated TPE. (c) Fully-stimulated TPE. (d) Energy diagram of fully-stimulated TPE in a direct-bandgap semiconductor with conduction band (CB) Cooper pair electrons recombining with valence band (VB) holes. (e) A schematic drawing of a superconducting light amplifier structure.

Our fully-quantum model shows broadband enhancement of both singly-stimulated (Fig. 1 b) and fully-stimulated (Fig. 1 c) Cooper-pair based TPE processes. The calculations are based on a full multimode treatment of the ultrafast light-matter interaction in a superconducting proximity region of the semiconductor, resulting in enhanced two-photon amplification. Besides enhancing two-photon interactions, the use of ultrafast pulses avoids any effects of light-induced heating and reduction of the superconducting order parameter. An intense ultrafast fs pulse may also reduce the order parameter locally by Cooper-pair breaking and heating [26,27,28,29], with two different time scales: the longer thermalization time and the shorter Cooper-pair breaking time.

The thermalization time is on the scale of several tens of ps, whereas the Cooper-pair breaking is on a shorter time scale of several ps. Despite being relatively short, this time is two orders of magnitude longer than typical ultrafast fs laser pulse duration. Moreover, it is worthy to note that the overall reduction of the order parameter is on the scale of few percent [28]. Therefore, the pulse can propagate through the amplifier, experiencing superconducting order parameter with standard ultrafast fs-pulse lasers operating at tens of MHz repetition rates. Our results show strong enhancement of the TPG close to the superconducting-gap related resonance. The bandwidth of the TPG is shown to become broader at higher carrier injection levels.

## 2. Theoretical model and results

In our proposal, we consider a direct-bandgap semiconductor vertical p-n junction with a top n-doped layer contacted by an *s*-wave superconductor (Fig. 1 e). For sufficiently thin n-doped layer, the p-n junction recombination region is in superconducting proximity [25,30]. We assume that the proximity itself has been established – as is usually assumed in the superconducting optoelectronics theory [1,2] and has been demonstrated experimentally [4]. The photonic structure in our proposal can be similar to that of edge-emitting semiconductor laser diodes and amplifiers [31]. The light pulse can propagate horizontally - perpendicular to the p-n junction and the current injection direction. Furthermore, we have considered multimode-photonic state, constituted of a single transversal wave-guide mode and broad range of longitudinal modes accounting for the short fs-scale pulses. In order to perform a full quantum-optical analysis of the ultrafast-pulse TPG, in our model the multimode photonic state is given in terms of coherent states on both inputs of the two-photon process $|Ph\rangle \propto \sum_{q_a,q_b}|\alpha_{q_a}\rangle|\alpha_{q_b}\rangle$. This state is waveguided in the proximity region p-n junction (Fig. 1 e), where the superconducting gap is in the semiconductor conduction band with electrons in a BCS state, while the valence band is in the normal state of holes (Fig. 1 d). The Fermi level of the superconductor contact attached to the n-type semiconductor is aligned with the conduction band of the semiconductor for effective electron Cooper-pair injection. Therefore at the p-n interface, the proximity is induced by Cooper-pair injection from the n-type contact side, and the radiative recombination is of normal holes and BCS correlated electrons [1]. By performing perturbative analysis we calculate the one-photon and two-photon stimulated emission spectra, and the corresponding TPG. Our model is based on a full BCS theory at finite temperatures, and it includes the BCS ground state, as well as the quasiparticle excitations out of the ground state, described by the Bogoliubov transformation:

$$c^{\dagger}_{\boldsymbol{k},\sigma}(t) = e^{i\tilde{\mu}_n t}\left(u_k e^{iE_k t}\gamma^{\dagger}_{\boldsymbol{k},\sigma} - s_\sigma v_k e^{-iE_k t}\gamma_{-\boldsymbol{k},\bar{\sigma}}\right) \tag{1}$$

where an electron creation operator $c^{\dagger}_{\boldsymbol{k},\sigma}$ is given in terms of the quasiparticle operators $\gamma_{\boldsymbol{k},\sigma}, \gamma^{\dagger}_{\boldsymbol{k},\sigma}$, $\tilde{\mu}_n = E_c + \frac{eV_a}{2} + \mu_n$, $E_k = \sqrt{\xi_n^2(k) + \Delta^2}$, $\xi_{n,p}(k) = \frac{k^2}{2m_{n,p}} - \mu_{n,p}$ , $\mu_n$ and $\mu_p$ are the electron quasi-Fermi level and hole quasi-Fermi level, respectively , $m_n$ and $m_p$ the electron and hole masses, respectively, $V_a$ the applied voltage, $E_c$ the edge of the conduction band, $2\Delta$ the superconducting gap, $s_\sigma = 1(-1)$ for $\sigma = \uparrow (\downarrow)$ and $u_k(v_k) = \{[1 + (-1)\xi_n(k)/E_k]/2\}^{1/2}$. The final state is the result of a light matter coupling Hamiltonian $H_I = \sum_{\boldsymbol{k},\boldsymbol{q},\sigma} B_{\boldsymbol{k},\boldsymbol{q}} b_{\boldsymbol{k}-\boldsymbol{q},\sigma} c_{\boldsymbol{k},\sigma} a_{\boldsymbol{q}}^{\dagger} + H.c.$ operating on the initial state.

In terms of the BCS quasiparticle operators the light-matter coupling Hamiltonian, using natural units ($\hbar = c = 1$) is:

$$H_I = \sum_{\boldsymbol{k},\sigma} \left( \sum_{\boldsymbol{q_1}} B_{\boldsymbol{k},\boldsymbol{q_1}} u_k \gamma_{\boldsymbol{k},\sigma} b_{\boldsymbol{k}-\boldsymbol{q_1},\sigma} a_{\boldsymbol{q_1}}^{\dagger} - s_\sigma \sum_{\boldsymbol{q_2}} B_{\boldsymbol{k},\boldsymbol{q_2}} v_k \gamma_{-\boldsymbol{k},\bar{\sigma}}^{\dagger} b_{\boldsymbol{k}-\boldsymbol{q_2},\sigma} a_{\boldsymbol{q_2}}^{\dagger} \right) + H.c. \tag{2}$$

where $B_{\boldsymbol{k},\boldsymbol{q}}$ is the coupling energy, $b_{\boldsymbol{k},\sigma}$ and $\gamma_{\boldsymbol{k},\sigma}$ are hole and BCS quasiparticle annihilation operators, respectively, with crystal momentum $\boldsymbol{k}$ and spin $\sigma$, and $a_{\boldsymbol{q}}^{\dagger}$ is the photon creation operator with momentum $\boldsymbol{q}$. The initial state of the system is given by $|\chi_i\rangle = |Ph\rangle|FS\rangle|BCS\rangle$, where $|Ph\rangle$ represents the multimode photonic coherent state, $|FS\rangle$ denotes the Fermi sea of holes in the valence band, and $|BCS\rangle$ is the electron superconducting BCS state. The two-photon interactions are represented by two-vertex Feynman diagrams (Fig. 1). The Green functions resulting from non-vanishing $c^{\dagger}c^{\dagger}$-type terms in the superconducting state are represented by double-arrowed electron propagators [32]. These Green functions allow pair emission through a single connected second-order Feynman diagram, contrary to the disconnected pair of first-order single-electron transitions. Using the interaction picture, the 1st and 2nd order contributions to the time evolution of the initial state are $|\chi_t(1)\rangle = -i\int_{-\infty}^{t} dt' H_I(t')|\chi_i\rangle$ and $|\chi_t(2)\rangle = -\int_{-\infty}^{t} dt' \int_{-\infty}^{t'} dt'' H_I(t') H_I(t'')|\chi_i\rangle$, respectively, where the interaction picture Hamiltonian is given by $H_I(t) = e^{iH_0 t} H_I e^{-iH_0 t}$ with $H_0$ the unperturbed Hamiltonian of BCS conduction band and a Fermi sea of holes in the valence band.

The transition amplitudes for the one-photon emission process and the TPE process are given by $A^{(1)} = \langle \chi_{f_e}(1) | \chi_t(1) \rangle = -i\int_{-\infty}^{t} dt' \langle \chi_{f_e}(1) | H(t') | \chi_i \rangle$ , $A^{(2)} = \langle \chi_{f_e}(2) | \chi_t(2) \rangle =$

$-\int_{-\infty}^{t} dt_3 \int_{-\infty}^{t_3} dt_4 \langle \chi_{f_e}(2) | H(t_3) H(t_4) | \chi_i \rangle$ respectively with $|\chi_{f_e}(1)\rangle = C_{1e} \sum_{\boldsymbol{k},\boldsymbol{q},\sigma} \left( u_k \gamma_{\boldsymbol{k},\sigma} b_{\boldsymbol{k}-\boldsymbol{q_1},\sigma} a_{\boldsymbol{q_1}}^{\dagger} - s_\sigma v_k \gamma_{-\boldsymbol{k},\bar{\sigma}}^{\dagger} b_{\boldsymbol{k}-\boldsymbol{q_2},\sigma} a_{\boldsymbol{q_2}}^{\dagger} \right) |Ph\rangle |BCS\rangle |FS\rangle$ and $|\chi_{f_e}(2)\rangle = C_e \sum_{\boldsymbol{k_1},\boldsymbol{k_2},\boldsymbol{p_1},\boldsymbol{q_1},\boldsymbol{q_2},\sigma_1,\sigma_2} \left( u_{k_2} \gamma_{\boldsymbol{k_2},\sigma_2} b_{\boldsymbol{p_2},\sigma_2} a_{\boldsymbol{q_2}}^{\dagger} - s_{\sigma_2} v_{k_2} \gamma_{-\boldsymbol{k_2},\bar{\sigma}_2}^{\dagger} b_{\boldsymbol{p_2},\sigma_2} a_{\boldsymbol{q_2}}^{\dagger} \right) \times \left( u_{k_1} \gamma_{\boldsymbol{k_1},\sigma_1} b_{\boldsymbol{p_1},\sigma_1} a_{\boldsymbol{q_1}}^{\dagger} - s_{\sigma_1} v_{k_1} \gamma_{-\boldsymbol{k_1},\bar{\sigma}_1}^{\dagger} b_{\boldsymbol{p_1},\sigma_1} a_{\boldsymbol{q_1}}^{\dagger} \right) |Ph\rangle |BCS\rangle |FS\rangle$. Assuming that the Fermi-Dirac distributions $f_k^{p(n)}$ and $u_k, v_k$ are slowly-varying on the scale of $q$, and neglecting the dependence of $B_k$ on $k$ ($B_k = B$), the one-photon emission rate is:

$$R_e^{(1)} = \frac{16\pi}{N^2} |C_{1e} B|^2 \sum_{\boldsymbol{k},\boldsymbol{q}} \delta(\Omega) \left(f_k^p\right)^2 \begin{bmatrix} |u_k|^4 \left(N + \left|\alpha_{q_1}\right|^2\right)^2 (f_k^n)^2 \\ + |v_k|^4 \left(N + \left|\alpha_{q_2}\right|^2\right)^2 (1 - f_k^n)^2 \end{bmatrix} \tag{3}$$

where $N$ is the number of photonic modes, $\Omega = \omega_q - \epsilon_p(\boldsymbol{k} - \boldsymbol{q}) - \tilde{\mu}_n = \omega_q - E_g - \mu_n - \mu_p - \xi_p(\boldsymbol{k} - \boldsymbol{q})$, $E_g = E_c + E_v + eV_a$, $E_v$ the edge of the conduction band, $\omega_q, \epsilon_p, \epsilon_n$ are photon, hole, and electron energies respectively, and $|C_{1e}|^2$ is the normalization constant. It is worthy to note that the normal one-photon behavior can be obtained by neglecting the effects of superconductivity. This is equivalent to taking the limit $\Delta \to 0$ for electron-like quasiparticles so that $u_k \to 1$, $v_k \to 0$ and $E_k \to \xi_n(k)$. This results in the normal-material based emission rate $R_e^{(1)} = \frac{16\pi}{N^2} |C_{1e} B|^2 \sum_{\boldsymbol{k},\boldsymbol{q}} \delta(\Omega) \left(N + \left|\alpha_{q_1}\right|^2\right)^2 \left(f_k^p\right)^2 (f_k^n)^2$. The explicit dependence of the rate on $\Delta$ is given essentially by $u_k, v_k$ where $u_k(v_k) = \left\{\left[1 + (-1)\xi_n(k)/\sqrt{\xi_n^2(k) + \Delta^2}\right]/2\right\}^{1/2}$. Next, approximating the summation over $k$ momenta by integration, using a 2D state density of a thin proximity layer, the one-photon rate is:

$$R_e^{(1)} = C_1 \sum_{\boldsymbol{q}} \left(f_{\xi_n}^p\right)^2 \frac{\Theta(\omega_{\boldsymbol{q}} - 2\tilde{\mu}_n + \mu_n)}{\sum_{\boldsymbol{q}} \left(N + \left|\alpha_{\boldsymbol{q}}\right|^2\right)} \begin{bmatrix} \left(1 + \Omega_q\right)^2 \left(N + \left|\alpha_{\boldsymbol{q_1}}\right|^2\right)^2 \left(f_{\xi_n}^n\right)^2 + \\ \left(1 - \Omega_q\right)^2 \left(N + \left|\alpha_{\boldsymbol{q_2}}\right|^2\right)^2 \left(1 - f_{\xi_n}^n\right)^2 \end{bmatrix} \tag{4}$$

where $\Omega_q = (\omega_{\boldsymbol{q}} - 2\tilde{\mu}_n)/\sqrt{(\omega_{\boldsymbol{q}} - 2\tilde{\mu}_n)^2 + \Delta^2}$ with $\Theta$ a step function, $\xi_n = \omega_{\boldsymbol{q}} - 2\tilde{\mu}_n$ , $\omega_{\boldsymbol{q}_{1/2}} = \omega_{\boldsymbol{q}} \pm \sqrt{(\omega_{\boldsymbol{q}} - 2\tilde{\mu}_n)^2 + \Delta^2}$ and $C_1$ a constant proportional to $|B|^2$. A similar procedure is carried out to calculate the one-photon absorption rate. The one-photon gain is given by $g = (v_g N_{ph})^{-1} dN_{ph}/dt$, where $v_g$ is the group velocity, $N_{ph} = |\alpha_q|^2$ is the average photon number in a coherent state $|\alpha_q\rangle$. The quantity $dN_{ph}/dt$ is the net photon emission rate which is given by $R_e - R_a$ where $R_e$ and $R_a$ are the photon emission and absorption rates, respectively. Therefore, the one-photon gain is:

$$g^{(1)} = \frac{C_1}{v_g} \sum_{\boldsymbol{q}} \frac{\Theta(\omega_{\boldsymbol{q}} - 2\tilde{\mu}_n + \mu_n)}{\sum_{\boldsymbol{q}} |\alpha_q|^2} \times \left[ \begin{array}{c} \frac{\left(f^p_{\xi_n}\right)^2}{\sum_{\boldsymbol{q}} \left(N + |\alpha_q|^2\right)} \left( \begin{array}{c} \left(N + |\alpha_{q_1}|^2\right)^2 (1 + \Omega_q)^2 (f^n_{\xi_n})^2 + \\ \left(N + |\alpha_{q_2}|^2\right)^2 (1 - \Omega_q)^2 (1 - f^n_{\xi_n})^2 \end{array} \right) \\ - \frac{\left(1 - f^p_{\xi_n}\right)^2}{\sum_{\boldsymbol{q}} |\alpha_q|^2} \left( \begin{array}{c} |\alpha_{q_1}|^4 (1 + \Omega_q)^2 (1 - f^n_{\xi_n})^2 \\ + |\alpha_{q_2}|^4 (1 - \Omega_q)^2 (1 - f^n_{\xi_n})^2 \end{array} \right) \end{array} \right] \tag{5}$$

This expression can be significantly simplified by taking the semiclassical approximation $|\alpha_q| \gg 1$ thus neglecting the spontaneous emission terms compared to the stimulated ones. To further simplify, in the limit of a single-mode case at zero temperature ($T = 0$) the gain is:

$$g^{(1)} = \frac{C_1}{v_g} \left[ \begin{array}{c} \Theta(2\tilde{\mu}_n - \omega_{\boldsymbol{q}}) \Theta(\omega_{\boldsymbol{q}} - 2\tilde{\mu}_n + \mu_n)(1 - \Omega_q)^2 \\ -\Theta(\omega_{\boldsymbol{q}} - 2\tilde{\mu}_n)(1 + \Omega_q)^2 \end{array} \right] \tag{6}$$

This one-photon gain, as expected, has no dependence on the seed intensity, and has a negligible dependence on $\Delta$ since $\Delta \ll 2\tilde{\mu}_n - \omega_{\boldsymbol{q}}$. The step functions define a lower cutoff $E_g$ and a width of $\mu_p + \sqrt{\mu_n^2 + \Delta^2}$.

A much richer behavior appears in the following calculation of the superconductivity-enhanced TPG. The matrix element for the stimulated TPE in this case is given by:

$$\langle \chi_{f_e}(2) | H(t_3) H(t_4) | \chi_i \rangle = \sum_{\boldsymbol{k_1} \dots \boldsymbol{k_4}, \boldsymbol{p_1}, \boldsymbol{q_1} \dots \boldsymbol{q_4}, \sigma_1 \dots \sigma_4} C_{2e}^{*} B_{\boldsymbol{k_3}, \boldsymbol{q_3}} B_{\boldsymbol{k_4}, \boldsymbol{q_4}} e^{i\Omega_3 t_3} e^{i\Omega_4 t_4} I_{Ph} I_{FS} I_{BCS} \qquad (7)$$

where $\Omega_i = \omega_{q_i} - \epsilon_p(\boldsymbol{k_i} - \boldsymbol{q_i}) - \tilde{\mu}_n = \omega_{q_i} - E_g - \mu_n - \mu_p - \xi_p(\boldsymbol{k_i} - \boldsymbol{q_i})$. We have distinguished, for the sake of clarity, between the different contributions of the various parts of the expression. The first term is the photonic term, given by:

$$I_{Ph} = \frac{1}{2N^2 - N} \sum_{\boldsymbol{q_a}, \boldsymbol{q_b}} \langle \alpha_{\boldsymbol{q_a}} | \langle \alpha_{\boldsymbol{q_b}} | \, a_{\boldsymbol{q_1}} a_{\boldsymbol{q_2}} a_{\boldsymbol{q_3}}^{\dagger} a_{\boldsymbol{q_4}}^{\dagger} \sum_{\boldsymbol{q_c}, \boldsymbol{q_d}} | \alpha_{\boldsymbol{q_c}} \rangle | \alpha_{\boldsymbol{q_d}} \rangle = Q_{SPT} + Q_{SST} + Q_{FST} \qquad (8)$$

with $Q_{SPT}$, $Q_{SST}$, $Q_{FST}$ representing the spontaneous, singly-stimulated and fully-stimulated TPE transitions, respectively (Fig. 1). More explicitly, these terms are:

$$Q_{SPT} = \left( \delta_{\boldsymbol{q_1}, \boldsymbol{q_3}} \delta_{\boldsymbol{q_2}, \boldsymbol{q_4}} + \delta_{\boldsymbol{q_2}, \boldsymbol{q_3}} \delta_{\boldsymbol{q_1}, \boldsymbol{q_4}} \right) \qquad (9)$$

$$Q_{SST} = \frac{4}{2N^2 - N} \begin{bmatrix} (N-1) Q_{SPT} \left( \left| \alpha_{\boldsymbol{q_1}} \right|^2 + \left| \alpha_{\boldsymbol{q_2}} \right|^2 \right) \\ + \left( \delta_{\boldsymbol{q_1}, \boldsymbol{q_3}} \alpha_{\boldsymbol{q_4}}^{*} \alpha_{\boldsymbol{q_2}} + \delta_{\boldsymbol{q_1}, \boldsymbol{q_4}} \alpha_{\boldsymbol{q_3}}^{*} \alpha_{\boldsymbol{q_2}} + \delta_{\boldsymbol{q_2}, \boldsymbol{q_3}} \alpha_{\boldsymbol{q_4}}^{*} \alpha_{\boldsymbol{q_1}} + \delta_{\boldsymbol{q_2}, \boldsymbol{q_4}} \alpha_{\boldsymbol{q_3}}^{*} \alpha_{\boldsymbol{q_1}} \right) \end{bmatrix} \qquad (10)$$

$$Q_{FST} = \frac{4}{2N^2 - N} \begin{bmatrix} \left( \delta_{\boldsymbol{q_1}, \boldsymbol{q_2}, \boldsymbol{q_3}} + \delta_{\boldsymbol{q_1}, \boldsymbol{q_2}, \boldsymbol{q_4}} + \delta_{\boldsymbol{q_1}, \boldsymbol{q_3}, \boldsymbol{q_4}} + \delta_{\boldsymbol{q_2}, \boldsymbol{q_3}, \boldsymbol{q_4}} \right) \alpha_{\boldsymbol{q_4}}^{*} \alpha_{\boldsymbol{q_3}}^{*} \alpha_{\boldsymbol{q_2}} \alpha_{\boldsymbol{q_1}} \\ \delta_{\boldsymbol{q_1}, \boldsymbol{q_2}} \delta_{\boldsymbol{q_3}, \boldsymbol{q_4}} \left( \alpha_{\boldsymbol{q_3}}^{*} \alpha_{\boldsymbol{q_1}} \right)^2 + \left( Q_{SPT} + (N-4) \delta_{\boldsymbol{q_1}, \boldsymbol{q_2}, \boldsymbol{q_3}, \boldsymbol{q_4}} \right) \left| \alpha_{\boldsymbol{q_1}} \right|^2 \left| \alpha_{\boldsymbol{q_2}} \right|^2 \end{bmatrix} \qquad (11)$$

As it can be seen from Eqs. (8-11), $Q_{SPT}$ does not depend on the seed intensity, $Q_{SST}$ is $\mathcal{O}(\alpha^2)$ – proportional to the seed intensity, and $Q_{FST}$ is $\mathcal{O}(\alpha^4)$ - proportional to the square of the seed intensity. A unique property of two-photon processes and TPG in particular, is the possibility of employing interference of various two-photon transition paths involving different spectral components of ultrafast pulses for coherent control of the process [33]. Our results clearly demonstrate the possibility of coherent control in superconducting two-photon gain based on such interference terms (Eqs. 10,11). Next, we calculate the valence band hole Fermi-sea, and conduction band electron BCS contributions:

$$I_{FS} = \langle FS | b^{\dagger}_{\boldsymbol{p_1},\sigma_1} b^{\dagger}_{\boldsymbol{p_2},\sigma_2} b_{\boldsymbol{k_3}-\boldsymbol{q_3},\sigma_3} b_{\boldsymbol{k_4}-\boldsymbol{q_4},\sigma_4} | FS \rangle$$
$$= f^p_{\boldsymbol{p_1}} f^p_{\boldsymbol{p_2}} \begin{pmatrix} \delta_{\boldsymbol{p_1},\boldsymbol{k_4}-\boldsymbol{q_4}} \delta_{\sigma_1,\sigma_4} \delta_{\boldsymbol{p_2},\boldsymbol{k_3}-\boldsymbol{q_3}} \delta_{\sigma_2,\sigma_3} \\ -\delta_{\boldsymbol{p_1},\boldsymbol{k_3}-\boldsymbol{q_3}} \delta_{\sigma_1,\sigma_3} \delta_{\boldsymbol{p_2},\boldsymbol{k_4}-\boldsymbol{q_4}} \delta_{\sigma_2,\sigma_4} \end{pmatrix} \tag{12}$$

$$I_{BCS} = \langle BCS | c^{\dagger}_{\boldsymbol{k_1},\sigma_1} c^{\dagger}_{\boldsymbol{k_2},\sigma_2} c_{\boldsymbol{k_3},\sigma_3}(t_3) c_{\boldsymbol{k_4},\sigma_4}(t_4) | BCS \rangle$$
$$= u_{k_1} v_{k_1} u_{k_3} v_{k_3} \delta_{\boldsymbol{k_1},-\boldsymbol{k_2}} \delta_{\sigma_1,\bar{\sigma}_2} \delta_{\boldsymbol{k}_3,-\boldsymbol{k_4}} \delta_{\sigma_3,\bar{\sigma}_4}$$
$$\times s_{\sigma_1} s_{\sigma_3} \begin{bmatrix} e^{iE_{k_3}(t_3-t_4)} \left( (1-f^n_{k_1}) f^n_{k_3} - f^n_{k_1} f^n_{k_3} \right) + \\ e^{-iE_{k_3}(t_3-t_4)} \left( f^n_{k_1}(1-f^n_{k_3}) - (1-f^n_{k_1})(1-f^n_{k_3}) \right) \end{bmatrix} \tag{13}$$

Thus, the TPE rate is:

$$R_e^{(2)} = \frac{8\pi}{2N^2 - N} |B|^4 \frac{1}{\sum_{\boldsymbol{p}} \left(f^p_p\right)^2} \sum_{\boldsymbol{k},\boldsymbol{q_1},\boldsymbol{q_2}} \frac{\Delta^2}{E_k^2} \frac{|C + I_{SST} + I_{FST}|^2 \left(f^p_{\xi_n}\right)^4}{\sum_{\boldsymbol{q_1},\boldsymbol{q_2}} [C + I_{SST} + I_{FST}]}$$
$$\times \delta\left(\tilde{\Omega}_2 + \tilde{\Omega}_1\right) \left[ \frac{(f^n_k)^2}{\left(\tilde{\Omega}_1 - E_k\right)^2} - 2\frac{f^n_k(1-f^n_k)}{\tilde{\Omega}_1^2 - E_k^2} + \frac{(1-f^n_k)^2}{\left(\tilde{\Omega}_1 + E_k\right)^2} \right] \tag{14}$$

with $I_{SST}, I_{FST}$ the singly- and fully-stimulated emission terms (Fig. 1 b,c), respectively, which are acquired from $Q_{SST}$ and $Q_{FST}$ after summation, $\tilde{\Omega}_1 = \omega_{q_i} - \epsilon_p(-\boldsymbol{k} - \boldsymbol{q_1}) - \tilde{\mu}_n$ and $\tilde{\Omega}_2 = \omega_{q_i} - \epsilon_p(\boldsymbol{k} - \boldsymbol{q_2}) - \tilde{\mu}_n$. Approximating the sum over the momenta by integration, the rate is:

$$R_e^{(2)} = C_2 |B|^4 \sum_{\boldsymbol{q_1},\boldsymbol{q_2}} \frac{|C + I_{SST} + I_{FST}|^2 \left(f^p_{\xi_n}\right)^4}{\sum_{\boldsymbol{q_1},\boldsymbol{q_2}} [C + I_{SST} + I_{FST}]} \left(\frac{\Delta}{\Omega}\right)^2$$
$$\times \left[ \frac{\left(f^n_{\xi_n}\right)^2}{(\Delta\omega_{12} - \Omega)^2} - 2\frac{f^n_{\xi_n}\left(1-f^n_{\xi_n}\right)}{\Delta\omega_{12}^2 - \Omega^2} + \frac{\left(1-f^n_{\xi_n}\right)^2}{(\Delta\omega_{12} + \Omega)^2} \right] \tag{15}$$

where $\Omega^2 = \left(\omega_{\boldsymbol{q_1}} + \omega_{\boldsymbol{q_2}} - 2\tilde{\mu}_n\right)^2 + 4\Delta^2$, $\xi_n = \frac{1}{2}\left(\omega_{\boldsymbol{q_1}} + \omega_{\boldsymbol{q_2}} - 2\tilde{\mu}_n\right)$, $\Delta\omega_{12} = \omega_{\boldsymbol{q_1}} - \omega_{\boldsymbol{q_2}}$, and $C, C_2$ constants. This Cooper-pair based stimulated TPE rate attains a resonance as $\Delta\omega_{12}$ approaches $\Omega$ (Fig. 2,3), it is proportional to $\Delta^2$ and hence vanishes for temperatures higher than the superconducting transition temperature ($T_c$) where $\Delta$ vanishes. A similar expression is evaluated for the absorption rate. Lastly, the TPG can be calculated from the stimulated TPE rate (Eq. 15) yielding:

$$g^{(2)} = \sum_{\boldsymbol{q_1},\boldsymbol{q_2}} \left(\frac{\Delta}{\Omega}\right)^2 \left[ \begin{array}{c} \dfrac{C_e|I^e|^2\left(f_{\xi_n}^n\right)^2\left(f_{\xi_n}^p\right)^4 - C_a|I^a|^2\left(1-f_{\xi_n}^n\right)^2\left(1-f_{\xi_n}^p\right)^4}{\left(\omega_{\boldsymbol{q_1}} - \omega_{\boldsymbol{q_2}} - \Omega\right)^2} \\ + \dfrac{C_e|I^e|^2\left(1-f_{\xi_n}^n\right)^2\left(f_{\xi_n}^p\right)^4 - C_a|I^a|^2\left(f_{\xi_n}^n\right)^2\left(1-f_{\xi_n}^p\right)^4}{\left(\omega_{\boldsymbol{q_1}} - \omega_{\boldsymbol{q_2}} + \Omega\right)^2} \\ -2\dfrac{C_e|I^e|^2\left(f_{\xi_n}^p\right)^4 - C_a|I^a|^2\left(1-f_{\xi_n}^p\right)^4}{\left(\omega_{\boldsymbol{q_1}} - \omega_{\boldsymbol{q_2}} - \Omega\right)\left(\omega_{\boldsymbol{q_1}} - \omega_{\boldsymbol{q_2}} + \Omega\right)} f_{\xi_n}^n\left(1-f_{\xi_n}^n\right) \end{array} \right] \tag{16}$$

where we have defined $I_e = C + I_{SST} + I_{FST}^e$ and $I^a$, an equivalent expression for the absorption rate. $C_e, C_a$ are constants. Considering for simplicity a two-mode case at $T = 0$ with $|\alpha_q| \gg 1$ :

$$g^{(2)} = \frac{C_2|B|^4 I_{FST}}{\left|\alpha_{\boldsymbol{q_2}}\right|^2 + \left|\alpha_{\boldsymbol{q_1}}\right|^2}\left(\frac{\Delta}{\Omega}\right)^2 \left[ \begin{array}{c} \dfrac{\Theta\left(2\tilde{\mu}_n - \omega_{\boldsymbol{q_1}} + \omega_{\boldsymbol{q_2}}\right) - \Theta\left(\omega_{\boldsymbol{q_1}} + \omega_{\boldsymbol{q_2}} - 2\tilde{\mu}_n\right)}{(\Delta\omega_{12} + \Omega)^2} \\ + \dfrac{\Theta\left(2\tilde{\mu}_n - \omega_{\boldsymbol{q_2}} + \omega_{\boldsymbol{q_1}}\right) - \Theta\left(\omega_{\boldsymbol{q_1}} + \omega_{\boldsymbol{q_2}} - 2\tilde{\mu}_n\right)}{(\Delta\omega_{12} - \Omega)^2} \end{array} \right] \tag{17}$$

This gain has a linear dependence on the seed intensity – similar to the case of semiconductor TPG without superconductivity [20,21], and has a very broad spectrum with finite gain from 0 to $E_g + \mu_n + \mu_p$, (for $T = 0$). The spectra for finite temperatures are shown in Fig 2.

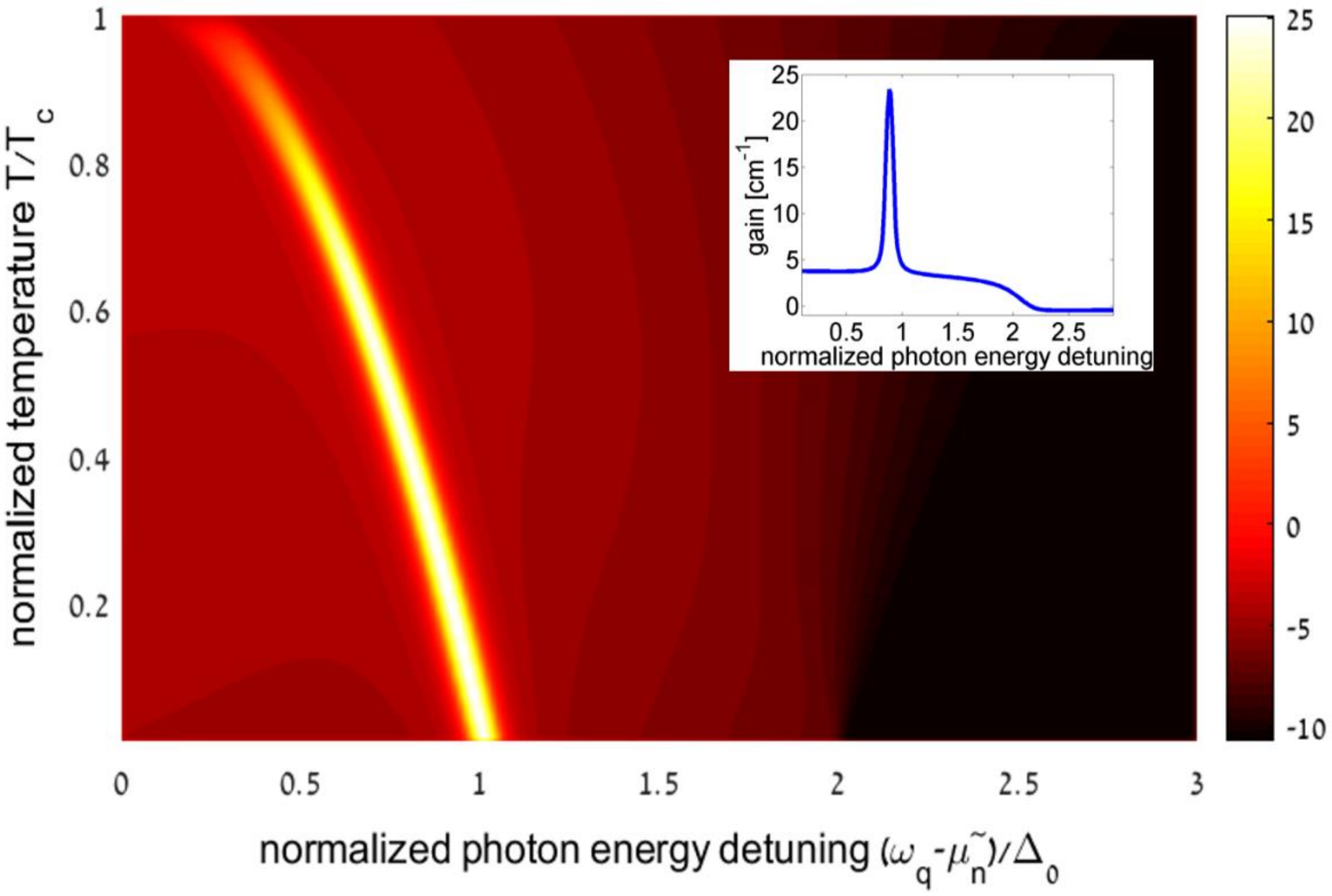

**Figure 2.** Cooper-pair based TPG spectra vs. temperature T, with $\Delta_0 \equiv \Delta(T = 0)$ for $|\alpha_q| \sim 400$. The gain is given in units of $cm^{-1}$. The inset shows a slice of the gain spectra at a given T.

We calculated the superconductivity-enhanced TPG, while taking into consideration the one-photon gain and loss under various conditions. The spectral shape of the gain has two characteristic bandwidth scales, one that is sharp and narrow pertaining to a resonance of the virtual state in the second order process as shown in Figure 1 (d), and another lesser in magnitude yet on a much broader scale. The spectral slice inset in Fig. 2 demonstrates this double-bandwidth shape of the gain spectra. Moreover, as the temperature approaches $T_c$, the effects of superconductivity, with the corresponding resonance at $\Delta$, disappear (Fig. 2). We also studied the effect of carrier injection level on the TPG. It can be seen in Fig. 3 that for increasing carrier density (quasi-Fermi level position) TPG increases and its spectrum becomes broader, with one-photon loss dominating the high-energy end of the spectrum.

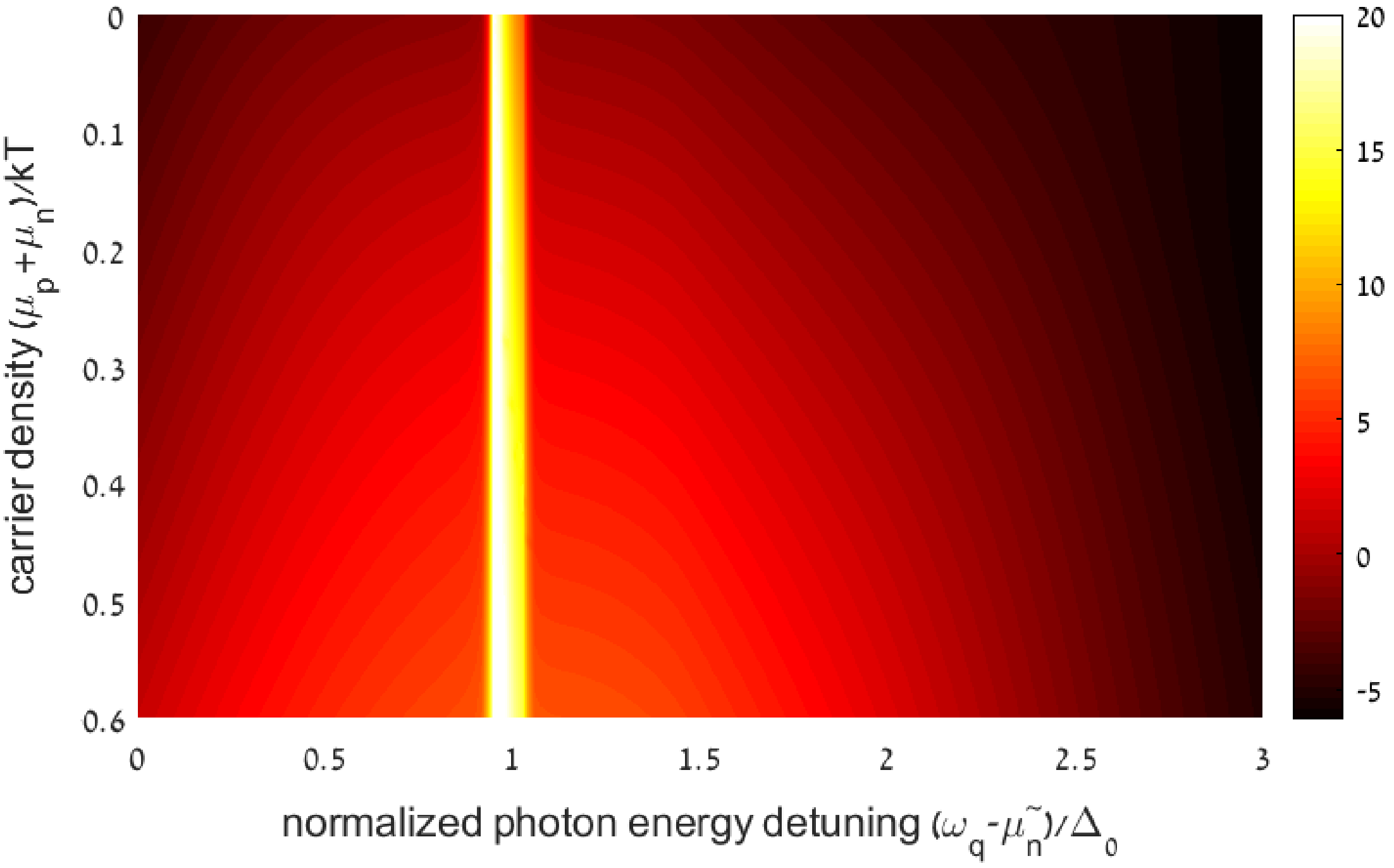

**Figure 3.** Cooper-pair based TPG spectra vs. quasi-Fermi level location (carrier density) for $|\alpha_q| \sim 400$. The gain is given in units of $cm^{-1}$.

In the low temperature limit $T \to 0$, and for a single-mode input with relatively high intensity $|\alpha_q| \gg 1$, a simple expression is obtained for the ratio between the one-photon and TPG:

$$\frac{g^{(2)}}{g^{(1)}} = \tilde{C}|B|^2|\alpha_q|^2\left(\frac{\Delta}{\Omega}\right)^2 \frac{[1-\Theta(2\omega_q - 2\tilde{\mu}_n)]}{[\Theta(2\tilde{\mu}_n - \omega_{\boldsymbol{q}})\Theta(\omega_{\boldsymbol{q}} - 2\tilde{\mu}_n + \mu_n) - \Theta(\omega_{\boldsymbol{q}} - 2\tilde{\mu}_n)]} \tag{18}$$

where $\tilde{C} = 4m_n\left(\mu_n - \Delta + \sqrt{\mu_n^2 + \Delta^2}\right)/m_p\mu_p\left[\left(\omega_{\boldsymbol{q}} - \tilde{\mu}_n\right)^2 + \Delta^2\right]$. This expression is proportional to the seed intensity and $\Delta^2$, and hence for moderately high intensities the superconductivity-enhanced TPG becomes comparable to its one-photon equivalent. Using characteristic values of III-V semiconductors we estimate that the TPG magnitude will approach its one-photon counterpart for $\left|\alpha_q\right|^2 \approx 10^7$ (Fig. 4), corresponding to pJ-scale pulse energy. For ultrafast pulses the bandwidth can introduce some variance to the pulse energy. Therefore, we estimate only the order of magnitude of the pulse energy

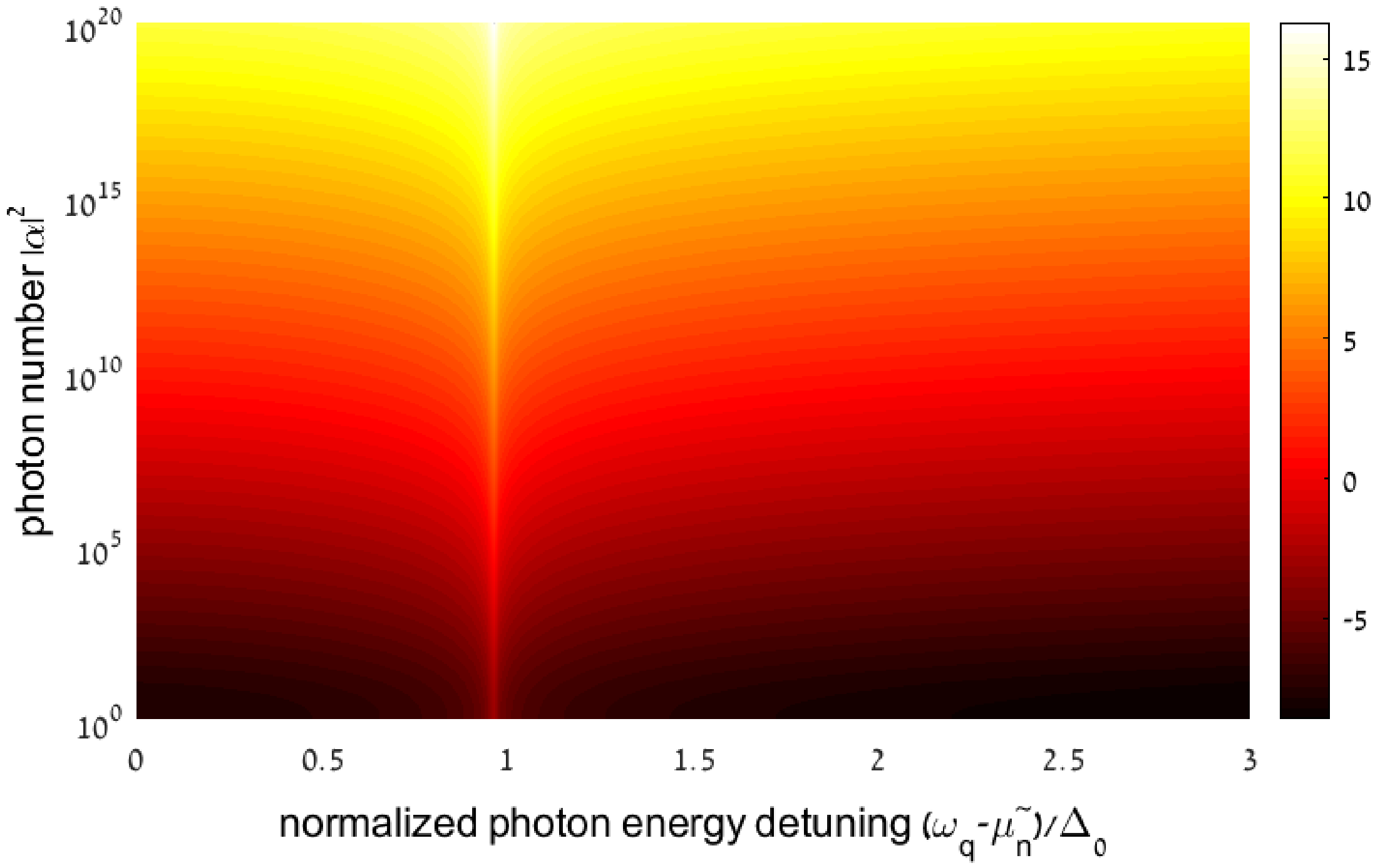

**Figure 4.** Cooper-pair based TPG to one-photon gain ratio spectra in log scale vs. photon number per pulse, $\left|\alpha_q\right|^2$ – for a typical III-V semiconductor.

This result is several orders of magnitude lower than the typical values of pulse energies required to observe TPG in semiconductors without superconductivity [21]. In principle in the extremely strong field limit nonlinear optical interactions with field strengths on the scale of ionization field strengths ( $E_0 \sim 5 \times 10^{11} Vm^{-1}$ [34]) could require non-perturbative modelling. However, in our calculations, the largest pulse energy considered is on the pJ scale, which corresponds to intensities about eight orders of magnitude smaller than ionization intensities

## 3. Conclusion

In conclusion, we have shown that Cooper-pair recombination in semiconductor-superconductor structures results in significant enhancement of ultrafast TPG. We developed a full quantum model of singly and fully-stimulated TPE in superconductivity enhanced recombination in semiconductors, allowing transition-path interference for coherent control applications. Furthermore, we have shown that at moderately high seed intensities the TPG contribution becomes comparable to the one-photon gain.

**References**

[1] Y. Asano, I. Suemune, H. Takayanagi, and E. Hanamura, "Luminescence of a Cooper-Pair" Phys. Rev. Lett. **103**, 187001 (2009).

[2] A. Hayat, H. Y. Kee, K. Burch, A. Steinberg, "Cooper-Pair-Based photon entanglement without isolated emitters" Phys. Rev. B **89**, 094508 (2014).

[3] M. Khoshnegar and A. H. Majedi, "Entangled photon pair generation in hybrid superconductor–semiconductor quantum dot devices", Phys. Rev. B **84**, 104504 (2011).

[4] I. Suemune, Y. Hayashi, S. Kuramitsu, K. Tanaka, T. Akazaki, H. Sasakura, R. Inoue, H. Takayanagi, Y. Asano, E. Hanamura, S. Odashima, and H. Kumano, "A Cooper-Pair Light-Emitting Diode: Temperature Dependence of Both Quantum Efficiency and Radiative Recombination Lifetime", Appl. Phys. Express **3**, 054001 (2010).

[5] V. Mourik, K. Zuo, S. M. Frolov, S. R. Plissard, E. P. A. M. Bakkers, and L. P. Kouwenhoven, "Signatures of Majorana fermions in hybrid superconductor-semiconductor nanowire devices" Science, **336**, 1003 (2012).

[6] S. De Franceschi, L. Kouwenhoven, C. Schönenberger, and W. Wernsdorfer, “Hybrid superconductor-quantum dot devices” Nature Nanotech. **5**, 703 (2010).

[7] F. P. Laussy , T. Taylor, I. A. Shelykh and A. V. Kavokin. "Superconductivity with excitons and polaritons: review and extension.", J. Nanophoton. **6**, 064502 (2012).

[8] A. Wallraff, D. I. Schuster, A. Blais, L. Frunzio, R.-S. Huang, J. Majer, S. Kumar, S. M. Girvin and R. J. Schoelkopf, "Circuit quantum electrodynamics: Coherent coupling of a single photon to a Cooper pair box", Nature **431**, 162 (2004).

[9] P. P. Sorokin and N. Braslau, "Some theoretical aspects of a proposed double quantum stimulated emission device", IBM J. Res. Dev. **8**, 177 (1964).

[10] A. M. Prokhorov, "Quantum electrodynamics", Science **149**, 828 (1965).

[11] E. del Valle, S. Zippilli, F. P. Laussy, A. Gonzalez-Tudela, G. Morigi, and C. Tejedor, "Two-photon lasing by a single quantum dot in a high-Q microcavity", Phys. Rev. B **81**, 035302 (2010).

[12] M. O. Scully, K. Wódkiewicz, M. S. Zubairy, J. Bergou, N. Lu, and J. Meyer ter Vehn, “Two-photon correlated-spontaneous-emission laser: Quantum noise quenching and squeezing” Phys. Rev. Lett. **60**, 1832 (1988).

[13] A. Hayat, P. Ginzburg, M. Orenstein, “Observation of two-photon emission from semiconductors,” Nature Photon. **2**, 238 (2008); A. Hayat , P. Ginzburg and M. Orenstein. "Measurement and model of the infrared two-photon emission spectrum of GaAs", Phys. Rev. Lett., **103**, 023601 (2009).

[14] Y. Ota, S Iwamoto, N. Kumagai, Y. Arakawa, "Spontaneous two-photon emission from a single quantum dot." Phys. Rev. Lett., **107**, 233602 (2011).

[15] H. Pattanaik, D. A. Fishman1 , S. Webster , D. J. Hagan and E. W. Van Stryland "Two-photon emission in direct-gap semiconductors." CIOMP-OSA Summer Session: Lasers and Their Applications. OSA, (2011).

[16] H. M. van Driel, "Semiconductor optics: On the path to entanglement." Nature Photon. **2**, 212 (2008).

[17] A. Hayat, P. Ginzburg, M. Orenstein., "High-rate entanglement source via two-photon emission from semiconductor quantum wells", Phys. Rev. B, **76**, 035339 (2007).

[18] Z. Lin and J. Vučković, "Enhanced two-photon processes in single quantum dots inside photonic crystal nanocavities", Phys. Rev. B **81**, 035301 (2010).

[19] A. N. Poddubny, P. Ginzburg, P. A. Belov, A. V. Zayats, and Y. S. Kivshar, "Tailoring and enhancing spontaneous two-photon emission using resonant plasmonic nanostructures", Phys. Rev. A **86**, 033826 (2012).

[20] C. N. Ironside, "Two-photon gain semiconductor amplifier", IEEE J. Quantum Electron. **28**, 842 (1992)

[21] A. Nevet, A. Hayat, M. Orenstein. "Measurement of optical two-photon gain in electrically pumped AlGaAs at room temperature", Phys. Rev. Lett. **104**, 207404 (2010).

[22] A. Nevet, A. Hayat, M. Orenstein, "Ultrafast pulse compression by semiconductor two-photon gain", Opt. Lett., **35**, 3877 (2010).

[23] D. C. Hutchings, E. W. Van Stryland, "Nondegenerate Two-Photon Absorption in Zinc Blende Semiconductors ," J. Opt. Soc. Am. B., **9**, 2065 (2011).

[24] M. Tinkham, "Introduction to Superconductivity", McGraw-Hill, New York, ed. 2, (1996).

[25] H. Sasakura, S. Kuramitsu, Y. Hayashi, K. Tanaka, T. Akazaki, E. Hanamura, R. Inoue, H. Takayanagi, Y. Asano, C. Hermannstädter, H. Kumano, and I. Suemune, "Enhanced Photon

Generation in a Nb/n-InGaAs/p-InP superconductor/semiconductor-diode light emitting device" Phys. Rev. Lett. **107**, 157403 (2011).

[26] M. Lindgren, M. Currie, W. S. Zeng, R. Sobolewski, S. Cherednichenko, B. Voronov and G. N. Gol'tsman, "Picosecond response of a superconducting hot-electron NbN photodetector" Applied Superconductivity **6**, 423-428 (1998).

[27] K. S. Il'in, A. A. Verevkin, G. N. Gol'tsman and R. Sobolewski, "Infrared hot-electron NbN superconducting photodetectors for imaging applications" Supercond. Sci. Technol. **12,** 755 (1999).

[28] M. Beck, M. Klammer, S. Lang, P. Leiderer, V. V. Kabanov, G. N. Gol'tsman, and J. Demsar " Energy-gap dynamics of superconducting NbN thin films studied by time-resolved terahertz spectroscopy", Phys. Rev. Lett. **107**, 177007 (2011).

[29] F. Marsili, M. J. Stevens, A. Kozorezov, et al., "Hotspot relaxation dynamics in a current carrying superconductor", arXiv:1506.03129 (2015).

[30] A. Kastalsky, A. W. Kleinsasser, L. H. Greene, R. Bhat, F. P. Milliken, and J. P. Harbison, "Observation of pair currents in superconductor-semiconductor contacts" Phys. Rev. Lett. **67**, 3026 (1991).

[31] L. A. Coldren, and S. W. Corzine, Diode Lasers and Photonic Integrated Circuits (Wiley, New York, 1995).

[32] A. A. Abrikosov, L. P. Gor'kov, and I. E. Dzyaloshinskii, *Methods of Quantum Field Theory in Statistical Physics* (Prentice Hall, Englewood Cliffs, NJ, 1963).

[33] D. Meshulach and Y. Silberberg, "Coherent quantum control of two-photon transitions by a femtosecond laser pulse" Nature **396**, 239 (1998).

[34] R. Boyd, Nonlinear Optics, 2nd ed. (Academic, New York). (2003).